\documentclass[twocolumn,pre,showpacs,amsmath]{revtex4}
\usepackage{epsfig,amsfonts,amsbsy}

\newcommand{\bv}[1]{{\boldsymbol #1}}

\newcommand{\ep}{\epsilon}
\newcommand{\lambdac}{\lambda_{\rm c}}
\newcommand{\rmR}{{\rm R}}
\newcommand{\rmS}{{\rm S}}
\newcommand{\rmI}{{\rm I}}

\newcommand{\der}[2]{\frac{d #1}{d  #2}}

\begin{document}
\title{Intrinsic Unpredictability of Epidemic Outbreaks on Networks}
\author{Junya Iwai${}^1$ and Shin-ichi Sasa${}^2$}
\affiliation {
 ${}^1$
Department of Basic Science, The University of Tokyo, Komaba,  Tokyo,
53-8902, Japan
\\
${}^2$Department of Physics, Kyoto University, Kyoto 606-8502, Japan
}
\date{\today}

\begin{abstract}
It has been known that epidemic outbreaks in the SIR model
on networks are described by phase transitions. Despite 
the similarity with percolation transitions, 
whether an epidemic outbreak occurs 
or not cannot be 
 predicted with probability one in the thermodynamic limit. 
We elucidate its  mechanism by deriving a simple 
Langevin equation that captures an essential aspect of the
phenomenon.  
We also calculate the probability of epidemic outbreaks 
near the transition point. 
\end{abstract}


\maketitle

\section{Introduction}


We start with 
the following 
 question: How can it be determined 
 whether an  epidemic outbreak 
 has occurred. 
Obviously, this 
 is hard to answer, because an accurate model of epidemic spread 
in real societies, which include complicated and heterogeneous human-to-human 
contact, cannot be constructed. Then, is it possible to predict the outbreak 
for a simple  mathematical model?  
Even in this case, the manner of the 
 early spread of  disease may
significantly influence states that manifest 
 after a sufficiently 
long time. For example, it seems reasonable to  conjecture that whether 
a single infected individual with a very high infection rate 
causes an outbreak 
may depend on the number of people infected by the individual,
which is essentially 
 stochastic. In the present paper, we attempt to 
formulate this conjecture.


Specifically, we study the stochastic SIR model as the simplest epidemic 
model, where an edge in the network represents a human-to-human contact 
and the infection rate $\lambda$ (the infection probability 
per unit time in  each edge) is a parameter of the SIR model
(see e.g. Ref. \cite{allen2008introduction} for an 
 introduction to 
 the stochastic SIR model; see also Refs. \cite{boccaletti2006complex,RevModPhys.81.591} 
for related social dynamics on complex networks). 
The SIR model may be defined for well-mixed cases 
\cite{bailey1950simple,bailey1953total,
metz1978epidemic,
martin1998final,PhysRevE.76.010901,
PhysRevE.86.062103},
homogeneous networks 
 \cite{diekmann1998deterministic,PhysRevE.64.050901,PhysRevE.66.016128,
lanvcic2011phase,bohman2012sir,moreno2002epidemic}, 
and scale-free networks 
\cite{moreno2002epidemic,PhysRevLett.86.3200,PhysRevE.64.066112,gallos2003distribution}. 
A remarkable phenomenon is that 
when $\lambda$ exceeds a critical value $\lambdac$, a disease 
spreads to macroscopic scales from a 
 single infected individual, 
which corresponds to an epidemic outbreak.  This was found in 
well-mixed cases  and random graphs, but $\lambdac=0$ for scale 
free networks. 
That is, epidemic outbreaks are described as  phase transition 
phenomena. In addition to the interest
in theoretical problems, recently, the SIR model on networks
has been studied so as to identify influential spreaders 
\cite{kitsak2010identification} and so as to determine 
 a better immunization strategy \cite{PhysRevLett.91.247901,
PhysRevLett.101.058701}. 


Although the phase transition in the SIR model may be a sort of 
percolation transition, its property is different from that of 
standard percolation models. In the SIR model exhibiting the
phase transition, the order parameter characterizing it may be 
the fraction of the infected population, 
 which is denoted by $\rho$.
Indeed, $\rho=0$ in the non-outbreak phase ($\lambda<\lambdac$), 
whereas 
 the expectation of $\rho$ becomes continuously non-zero 
from $0$ when $\lambda > \lambdac$. 
This phenomenon is in accordance with the standard percolation 
transition. However, on one hand, the order parameter in the percolated
phase, e.g. the fraction  of the largest  cluster, takes a definite value 
with probability one in the thermodynamic limit; on the other hand,  
the fraction of the 
 infected population in the SIR model is 
not uniquely determined even in the thermodynamic limit. In fact, 
 it has been reported that the distribution function of the order 
parameter in SIR models with finite sizes  shows two peaks at 
$\rho=0$ and $\rho=\rho_*$ for well-mixed cases 
\cite{bailey1953total,metz1978epidemic,martin1998final,PhysRevE.76.010901},
homogeneous networks \cite{diekmann1998deterministic,lanvcic2011phase,PhysRevE.64.050901}, 
and scale-free networks \cite{gallos2003distribution}. 
Mathematically, the probability density of $\rho$ in the thermodynamic 
limit may be expressed as 
\begin{equation}
P(\rho;\lambda)= (1-q(\lambda)) \delta(\rho)
+q(\lambda) \delta(\rho-\rho_*),
\label{goal}
\end{equation}
where $q=0$ for $\lambda \le \lambdac$ and 
$q\not = 0$ for $\lambda > \lambdac$. This means that 
the value of  
the fraction of the 
infected population 
 in the outbreak phase, 
which is either $0$ or $\rho_*(\lambda)$, cannot be predicted with certainty. 
We call this phenomenon the 
{\it intrinsic unpredictability of epidemic outbreaks}.  


In this paper, we  clearify the meaning 
 of \eqref{goal}.
We first observe the phenomenon in the SIR model defined on 
a random regular graph. By employing a mean field approximation,
we describe the epidemic spread dynamics in terms of a master 
equation for two variables.  Then, with a system size 
expansion, we approximate the solutions to the master equation  
by those to a Langevin equation. Now we can analyze this Langevin 
equation and work out 
 the mechanism of the appearance of 
the two peaks. We also calculate $q(\lambda)$ near the transition point.

\section{Model}\label{sec_model}               %


Let $G$ be  a random $k$-regular graph consisting of $N$ nodes.
For each $x \in G$, the state $\sigma(x) \in \{ \rmS, \rmI, \rmR \} $ 
is defined, where $\rmS$, $\rmI$, and $\rmR$ represent Susceptible, 
Infective, and Recovered, respectively.
The state of the whole system is given by 
 $(\sigma_x)_{x \in G}$,
which is denoted by  $\bv{\sigma}$ collectively.
The SIR model on networks is described by 
a continuous time Markov process 
with 
 infection rate $\lambda$ and recovery rate $\mu$. 
Concretely, the transition rate $W(\bv{\sigma} \to \bv{\sigma}')$ 
of the Markov process is given as 
\begin{equation}
W(\bv{\sigma} \to \bv{\sigma}') 
= \sum_{x \in G} w(\bv{\sigma} \to \bv{\sigma}'|x),  
\end{equation}
with 
\begin{eqnarray}
w(\bv{\sigma} \to \bv{\sigma}'|x)& = & \lambda 
 \left[\delta(\sigma_x,\rmS)\delta(\sigma_x',\rmI)
\sum_{y \in B(x)} \delta(\sigma_y,\rmI) \right]  \nonumber \\
& & +\mu \delta(\sigma_x, \rmI)\delta(\sigma_x', \rmR), 
\label{rate-netmodel}
\end{eqnarray}
where $B(x)$ is a set of $k$-adjacent nodes to $x \in G$.
Hereinafter, 
 without loss of generality, we use dimensionless 
time by setting $\mu=1$. For almost all time sequences, 
infective nodes  vanish after a sufficiently long time, 
and then the system reaches  a  stationary state, 
which is called the {\it final state}. 
The ratio of the total number of recovered nodes 
to $N$ in the final state is equivalent to the 
fraction of the 
infected population $\rho$. This quantity
measures the extent of the epidemic spread. At $t=0$, 
we assume that $\sigma=\rmI$ for only one node 
selected randomly and that $\sigma=\rmS$ for the other
nodes. 

\begin{figure}
\includegraphics[, width=0.8\columnwidth]{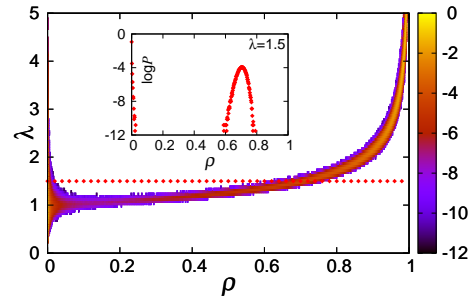}
\caption{(Color online) 
Color representation of $\log P(\rho;\lambda)$ in the $(\rho,
\lambda)$ plane. It is  obtained by numerical simulations of 
the SIR model  on a random regular graph. The inset shows 
$\log P(\rho;\lambda)$ as a function of $\rho$ for $\lambda=1.5$. 
}
\label{fig-sir-pm3d}
\end{figure}
\begin{figure}
\includegraphics[, width=0.8\columnwidth]{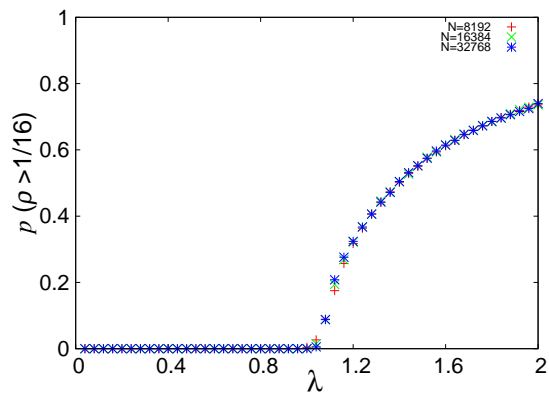}
\caption{
(Color online)
$p(\rho >1/16)$ as a function of $\lambda$ for several values of $N$.
}
\label{fig-sir-Nxxx-p0-N16}
\end{figure}

In Fig. \ref{fig-sir-pm3d}, as an example, we show  the result 
 of numerical simulations for the model with  $k=3$ and  $N=8192$. 
We measured the probability density  $P(\rho;\lambda)$ of the 
fraction of the 
infected population $\rho$ for various values of 
$\lambda$. 
This figure suggests that the expectation 
 of $\rho$ 
becomes  non-zero when $\lambda$ exceeds a critical value.
The important observation here is that $\log P $ in the outbreak
phase  has a sharp peak near $\rho=0$, too. Indeed, the inset 
in Fig. \ref{fig-sir-pm3d} clearly 
 shows the existence of the 
two peaks in  $\log P$ with $\lambda=1.5$. Similar graphs  were 
reported in Refs. \cite{bailey1953total,martin1998final,
gallos2003distribution,PhysRevE.76.010901,
PhysRevE.64.050901,lanvcic2011phase}. 
The existence
of the two peaks is not due to a finite size effect, as shown 
in Fig. \ref{fig-sir-Nxxx-p0-N16}, where the probability that 
$\rho > 1/16$, which is denoted by $p(\rho >1/16)$, is 
plotted as a function of $\lambda$ for several values of $N$. 
Note that $\lim_{N \to \infty} p(\rho >1/16)=q(\lambda)$ when
$\rho_*(\lambda) >1/16$. 
These results suggest the limiting density  (\ref{goal}), 
where $q(\lambda)$ becomes 
continuously non-zero for $\lambda > \lambdac$ whereas 
 $q(\lambda)=0$ 
for $\lambda \le \lambdac$. This is the phenomenon that we 
attempt to understand in this paper.

  
\section{Analysis}     %


Defining two variables $s\equiv \sum_{x}\delta(\sigma_x,S)/N$ 
and $i \equiv \sum_{x}\delta(\sigma_x,I)/N$, we consider 
a continuous-time Markov process of the two variables 
as an approximation of the SIR model on the network 
\cite{hufnagel2004forecast,colizza2006modeling}. 
We expect 
 the phenomenon we are concerned with to be 
reproduced within this approximation; we verify this at a later stage. 
The transition rate of $(s,i)\to (s,i-1/N)$ is exactly 
given as $Ni$, and we approximate the rate $(s,i)\to (s-1/N,i+1/N)$ 
as  $\lambda k N s \psi $, where $\psi$ is the probability 
of finding $y \in B(x)$ such that $\sigma_y=\rmI$ for any  $x$.
Here, the infective nodes 
 form a connected cluster,
and this cluster is tree-like because the 
 typical size of the 
 loops is $O(\log N)$. Now, as an approximation, we assume 
that there are $N i (k-2)$ edges connecting the tree-like 
cluster with susceptible nodes 
\cite{derrida1986random,keeling2005networks}. Therefore, 
$\psi$ is estimated 
as the rate of $N i (k-2)$ to the number of all edges $N k$
in the thermodynamic limit. That is, $\psi=i (k-2)/k$. 
Below, we focus on the case $k=3$.


Let $P(s,i,t)$ be the probability density of $s(t)=s$ and $i(t)=i$.
Then, $P(s,i,t)$ obeys the master equation
\begin{eqnarray}
\frac{\partial P(s,i,t)}{\partial t} 
&=& N \left( i+\frac{1}{N} \right) P \left(s,i+\frac{1}{N},t \right) 
-N i P \left(s,i,t \right) \nonumber \\
&+& N \lambda  \left (s+\frac{1}{N}\right) \left(i-\frac{1}{N}\right) 
P\left( s+\frac{1}{N},i-\frac{1}{N},t\right) \nonumber \\
&-& N\lambda  s i P\left(s,i,t\right).  
\label{MST}
\end{eqnarray}
When $N$ is sufficiently large, the master equation for $P(s,i,t)$
can be expanded as
\begin{equation}
\frac{\partial P}{\partial t}
+\partial_i J_i +\partial_s J_s+O\left( \frac{1}{N^2} \right)=0,
\label{sNMST}
\end{equation}
with 
\begin{eqnarray}
J_i&=& 
\left(\lambda s-1\right) i P 
- \partial_i\left[ \frac{\left(\lambda s+1\right)i}{2N} P \right] 
+ \partial_s \left(\frac{\lambda s i }{2N}P \right), \nonumber \\
J_s&=& 
-\lambda s i P 
- \partial_s \left( \frac{\lambda s i }{2N}P\right)
+\partial_i \left(\frac{\lambda s i}{2N} P \right).
\end{eqnarray}
By assuming that  $O(1/N^2)$ terms can be ignored, we 
obtain the Fokker-Planck equation \cite{gardiner2004handbook}.

\begin{figure}
\includegraphics[, width=0.8\columnwidth]{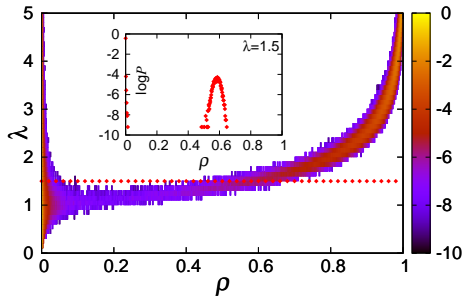}
\caption{
(Color online) Result of numerical simulations of 
\eqref{s_lgv} and  \eqref{i_lgv}. The presentation is 
the same as those in Fig. \ref{fig-sir-Nxxx-p0-N16}. $N=1024$ for 
the main frame and $N=8192$ for the inset.}
\label{fig-sy_lgv-pm3d}
\end{figure}

It can be confirmed by direct calculation that 
this Fokker-Planck equation (\ref{sNMST}) describes
the time evolution of the probability density for 
the following set of Langevin equations:
\begin{eqnarray}
\der{s}{t} &=& -\lambda s i -\sqrt{\frac{\lambda s i}{N}}\cdot \xi_1, 
\label{s_lgv} 
\\ 
\der{i}{t} &=& \lambda s i - i +\sqrt{\frac{\lambda s i}{N}}\cdot \xi_1
	+\sqrt{\frac{i}{N}}\cdot \xi_2, 
\label{i_lgv} 
\end{eqnarray}
where $\xi_i$ is  Gaussian white noise that satisfies 
$\left<\xi_i\left(t\right) \right>=0$
and $ \left<\xi_i\left(t\right) \xi_j\left(t'\right) \right>
=\delta_{i j}\delta\left(t-t'\right)$.
The symbol ``$\cdot$'' in front of $\xi_1$ and $\xi_2$ 
in \eqref{s_lgv} and \eqref{i_lgv} represents the 
Ito product rule. 
The same equations as 
\eqref{s_lgv} and \eqref{i_lgv} were presented in Refs. 
\cite{hufnagel2004forecast,colizza2006modeling}.
In this description, the fraction of the 
 infected population
is given by 
\begin{equation}
\rho=1-s(\infty).
\end{equation}
In Fig. \ref{fig-sy_lgv-pm3d}, we show the 
 result of numerical simulations of the Langevin equations \eqref{s_lgv}
and \eqref{i_lgv}. 
Comparing Fig. \ref{fig-sy_lgv-pm3d} with Fig. \ref{fig-sir-pm3d},
we find that the phenomenon under study 
 is described
by the Langevin equations \eqref{s_lgv} and \eqref{i_lgv}. 
Thus, our problem may be solved by analyzing them.


Now, the key idea of our analysis is the introduction 
of a new variable $Y=\sqrt{i N}$. Then, 
\eqref{s_lgv} and \eqref{i_lgv} are re-written as 
\begin{eqnarray}
\der{s}{t} &=&
\frac{1}{N}\left[ - \lambda s Y^2 -\sqrt{{\lambda s Y^2}}\cdot \xi_1\right],
\label{s2_lgv} \\
\der{Y}{t} &=& \frac{1}{2}\left\{\left(\lambda s - 1\right)Y
	 -\frac{1}{4}\left({\lambda s +1} \right)\frac{1}{Y}\right\} 
\nonumber 
\\
  && +\frac{1}{2}\sqrt{{\lambda s }}\cdot \xi_1 
      +\frac{1}{2}\sqrt{1}\cdot \xi_2, 
\label{Y_lgv}
\end{eqnarray}
where it should be noted that the multiplication 
of the variable $Y$ and the noise does not appear 
in \eqref{Y_lgv}.  
We then consider the probability $q(\lambda)$ in the 
thermodynamic limit as the probability 
of observing $Y \simeq N^{1/2}$, because it is 
equivalent to $\rho >0$.


Here, from \eqref{s2_lgv} and \eqref{Y_lgv}, 
we find that the characteristic 
time scale of $s$ is $N$ times that of $Y$. Thus, when 
$N$ is sufficiently large, $s$ almost retains 
 its value when $Y$ changes over time. 
  In particular, it is reasonable 
to set $s=1$ when $t$ is shorter than $N$.  
In this time interval, \eqref{Y_lgv} is expressed as
\begin{eqnarray}
\der{Y}{t} &=& -\partial_Y U(Y)+\sqrt{2D}\xi, 
\label{Y2_lgv}
\end{eqnarray}
where $D=(\lambda+1)/8$ and the potential $U(Y)$ is 
calculated as 
\begin{equation}
U(Y)=-\frac{1}{4}\left(\lambda  - 1\right)Y^2 
+\frac{1}{8}\left({\lambda +1}\right)\log(Y). 
\end{equation}
$\xi$ is  Gaussian white noise with unit variance,
where we have used the relation 
$\sqrt{\lambda}/2 \xi_1+1/2 \xi_2=\sqrt{\lambda+1}/2 \xi$.
The initial condition is given as $Y(0)=1$. 
It should be noted that  \eqref{Y2_lgv} is independent of $N$. 
Thus, solutions satisfying $Y\simeq N^{1/2}$ in \eqref{s2_lgv}
and \eqref{Y_lgv} correspond 
 to solutions satisfying $Y \to \infty$  
in \eqref{Y2_lgv}. We identify $q(\lambda)$ with  the probability of 
finding these solutions. We now 
 derive this probability.


\begin{figure}
\includegraphics[, width=0.8\columnwidth]{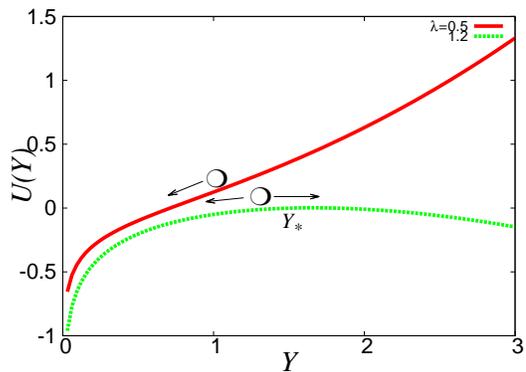}
\caption{
(Color online) Functional forms of $U(Y)$. 
$\lambda=0.5$ (red solid line) and $\lambda=1.2$ (green dotted line.)
}
\label{fig-U_Y}
\end{figure}



First, we investigate the shapes 
 of the graph
$U\left(Y\right)$. We find that 
$U\left(0_+\right)=-\infty$ for any $\lambda$ 
and that $U\left(Y\right)$ monotonically increases in $Y$ 
for $\lambda <1$, while  $U(Y)$ has a single maximum peak 
at $Y=Y_*$ for $\lambda >1$, 
where 
\begin{equation}
Y_\ast=\frac{1}{2}
\sqrt{\frac{\lambda+1}{\lambda-1}}. 
\end{equation}
As a reference, in Fig. \ref{fig-U_Y}, we show the shapes of 
$U\left(Y\right)$ for $\lambda=0.5$ and $1.2$. 


Next, based on the shapes  of the potential function, we discuss 
the 
expected behavior of solutions to \eqref{Y2_lgv}. When  
$\lambda<1$, the probability  of $Y \to \infty$ is 
obviously zero because  $U(Y)$ is a monotonically 
increasing function in $Y$. That is, $q(\lambda)=0$ 
in this case. 
The behavior for $\lambda>1$ is complicated. 
We thus focus on the case that $\lambda =1+\ep$, where
$\ep$ is a small positive number. In this case, 
$Y_*\simeq \ep^{-1/2}$. 
We then note 
 that if a  solution $Y$ to \eqref{Y2_lgv} happens 
to exceed $Y_*$, it is comparatively likely that 
$ Y \to \infty$. Assuming that the probability of $ Y \to \infty$
under the condition $Y \ge Y_*$ at some time is unity,
we estimate $q(\lambda)$ as the probability that $Y$ exceeds
$Y_*$. Furthermore, we express $q(\lambda)$ in 
terms of the transition rate $T$ from $Y=1$ to $Y=Y_*$. Noting that 
the transition rate from $Y=1$ to $Y=0$ is equal to
the recovery rate 
 in the original SIR model, we 
can write  
\begin{equation}
q=\frac{T}{1+T}.
\label{qT}
\end{equation}
Since $T$ is positive and finite,
we obtain  $0 < q(\lambda) < 1$.
In this manner, we have clearly explained the probabilistic 
nature in the outbreak phase, 
and we have obtained $\lambdac=1$. 

\begin{figure}
\includegraphics[, width=0.8\columnwidth]{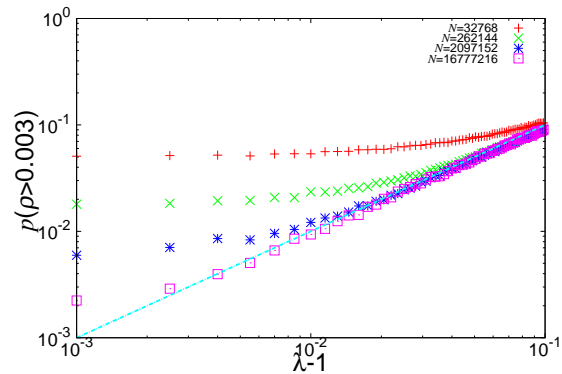}
\caption{
(Color online) 
$p(\rho >0.003)$ as a function of $\ep=\lambda-1$ 
obtained by numerical simulations of (\ref{s2_lgv})
and (\ref{Y_lgv}). The guide line represents 
$p(\rho> 0.003)=\ep$, which is expected from 
the theoretical analysis.
}
\label{fig-sy_lgv-sp-Nxxx-fit}
\end{figure}

Finally, we calculate $q(\lambda)$ quantitatively
near the transition point. From 
$Y_* \simeq \ep^{-1/2}$ and $U(Y_*) \simeq \log\ep$,
we estimate the slope of the straight line connecting two points
$\left(1,U(1)\right)$ and $\left(Y_*,U(Y_*)\right)$ in 
the $(Y,U)$ plane as  $(U(Y_*)-U(1))/(Y_*-1)\simeq 
\sqrt{\ep}(\log \ep)$, which approaches zero in the limit
$\ep\rightarrow 0$. Thus, 
the transition from $Y=1$ to $Y=Y_*$ may be assumed 
to be free Brownian motion with the diffusion constant 
$D=(\lambda+1)/8$. The transition rate from $Y=1$
to $Y_*$ is then estimated as 
$T=2D/Y_*^2  = \ep+ O(\ep^2)$.  We thus obtain
\begin{eqnarray}
q(\lambda) &=& \ep+O(\ep^2).
\label{qep}
\end{eqnarray}
In Fig. \ref{fig-sy_lgv-sp-Nxxx-fit}, we compare 
the theoretical result with those obtained in 
numerical simulations of \eqref{s2_lgv} and 
\eqref{Y_lgv}. We measured the probability that
$\rho > 0.003$, which is denoted as 
$p(\rho > 0.003)$.  Recall that 
$\lim_{N \to \infty} p(\rho >0.003) =q(\lambda)$
when $\rho_*(\lambda) > 0.003$.
Since the experimental result 
suggests $p(\rho >0.003)=\ep+O(\ep^2)$ in the limit 
$N \to \infty$,  we claim that the theoretical result 
(\ref{qep}) is in good agreement with the 
experimental result.


\section{Concluding remarks}               %


In this paper, we have achieved a novel understanding of 
the intrinsic unpredictability of epidemic outbreaks
by analyzing the Langevin equation \eqref{Y2_lgv},
 which 
 effectively describes this 
 singular phenomenon. 
Further, 
 trajectories in the outbreak 
phase are divided into two groups: 
 trajectories in one group are absorbed into zero, 
 and the others diverge
in \eqref{Y2_lgv}. 
The division  corresponds to the non-trivial limiting
density given in \eqref{goal}. 
On the basis of this description, we calculated 
the probability of an epidemic outbreak near 
the transition point. Before ending the paper, 
we make a few remarks. 


First, the probability $q(\lambda)$ was studied
in the 
 mathematical literature 
 (see \cite{yan2008distribution} 
and \cite{britton2010stochastic} as reviews.)  
To the best of our knowledge, the method proposed
in this paper has never been 
 used in previous studies. 
It might be interesting to connect our analysis 
with mathematical studies.

\begin{figure}
\includegraphics[, width=.8\columnwidth]{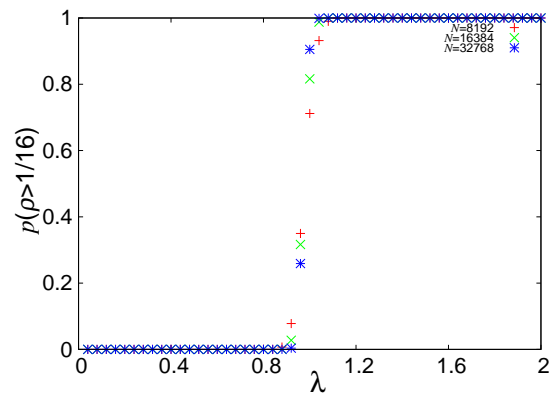}
\vspace{5mm}\caption{
Probability that  $\rho > 1/16$ in the SIR model on a random 
regular graph; $m=N/128$. 
}
\label{fig-sir-Nxxx-p0-mN128-N16}
\end{figure}

Second, although we have investigated the simplest model in this 
paper, similar analysis might be applied to various models. 
For example, we can consider the 
case that there are $m$ infected nodes at time 
$t=0$. Since the essence of the phenomenon is the 
existence of $Y_*$, the same result is obtained 
when $m$ is independent of $N$. 
However, for the case $m=c N$ with a small
positive number $c$,  $Y(t)$ is  never 
adsorbed to zero 
 in the outbreak phase,
because $Y(0)$ is 
infinitely far away from $Y=Y_*$. This
is qualitatively different from the case
$m=1$, which was reported in Refs. 
\cite{barbour1974functional, miller2012epidemics}.
In fact, as suggested in  Fig. \ref{fig-sir-Nxxx-p0-mN128-N16}, 
$q(\lambda)$ jumps discontinuously to $q(\lambda)=1$ 
which is similar to the behavior 
 observed in standard percolation transitions.


Finally, as another generalization, one may study the behavior of 
the SIR on more complex networks. In these cases, since the mean 
field approximation might not be effective, one needs to devise 
 a new technique to describe the unpredictability
of outbreaks.
Moreover, one of the most interesting is to predict
probabilistic epidemic outbreaks from limited data on realistic 
networks. We hope that future studies will address 
 these problems. 



The authors thank N. Nakagawa, T. Nemoto and M. Itami for their 
useful comments.
The present study was supported by KAKENHI No. 22340109 and 
No. 23654130.



\end{document}